%% file: main.tex
\documentclass[acmsmall, authorversion, nonacm]{acmart}
\usepackage{comment}
\usepackage{makecell}
\usepackage{rotating}
\usepackage{longtable}
\usepackage{adjustbox}
\usepackage[position =top]{subcaption}

\AtBeginDocument{%
  \providecommand\BibTeX{{%
    \normalfont B\kern-0.5em{\scshape i\kern-0.25em b}\kern-0.8em\TeX}}}

\acmDOI{10.2196/21212}

\begin{document}

%%
%% The "title" command has an optional parameter,
%% allowing the author to define a "short title" to be used in page headers.
\title{Patterns of Routes of Administration and Drug Tampering for Nonmedical Opioid Consumption: Data Mining and Content Analysis of Reddit Discussions}

%%
%% The "author" command and its associated commands are used to define
%% the authors and their affiliations.
%% Of note is the shared affiliation of the first two authors, and the
%% "authornote" and "authornotemark" commands
%% used to denote shared contribution to the research.
\author{Duilio Balsamo}
%\authornote{Both authors contributed equally to this research.}
\email{duilio.balsamo@unito.it}
\orcid{1234-5678-9012}
\affiliation{%
 \institution{Mathematics Department, University of Turin}
 \city{Turin}
 \state{Italy}
}

\author{Paolo Bajardi}
%\authornotemark[1]
\email{paolo.bajardi@isi.it}
\affiliation{%
 \institution{ISI Foundation}
 %\streetaddress{P.O. Box 1212}
 \city{Turin}
 \state{Italy}
%   \postcode{43017-6221}
}

\author{Alberto Salomone}
\email{alberto.salomone@unito.it}
\affiliation{
 \institution{Chemistry Department, University of Turin}
 \city{Turin}
 \state{Italy}
}

\author{Rossano Schifanella}
\email{rossano.schifanella@unito.it}
\affiliation{%
 \institution{Computer Science Department, University of Turin}
 \city{Turin}
 \state{Italy}
}
\affiliation{\institution{ISI Foundation}
 \city{Turin}
 \state{Italy}
 }

%%
%% By default, the full list of authors will be used in the page
%% headers. Often, this list is too long, and will overlap
%% other information printed in the page headers. This command allows
%% the author to define a more concise list
%% of authors' names for this purpose.

%\renewcommand{\shortauthors}{Balsamo et al.}

%%
%% The abstract is a short summary of the work to be presented in the
%% article.
\begin{abstract}
\textbf{Background:}
The complex unfolding of the US opioid epidemic in the last 20 years has been the subject of a large body of medical and pharmacological research, and it has sparked a multidisciplinary discussion on how to implement interventions and policies to effectively control its impact on public health.\\
\textbf{Objectives:}
This study leverages Reddit, a social media platform, as the primary data source to investigate the opioid crisis. We aimed to find a large cohort of Reddit users interested in discussing the use of opioids, trace the temporal evolution of their interest, and extensively characterize patterns of the nonmedical consumption of opioids, with a focus on routes of administration and drug tampering.\\
\textbf{Methods:}
We used a semiautomatic information retrieval algorithm to identify subreddits discussing nonmedical opioid consumption and developed a methodology based on word embedding to find alternative colloquial and nonmedical terms referring to opioid substances, routes of administration, and drug-tampering methods. We modeled the preferences of adoption of substances and routes of administration, estimating their prevalence and temporal unfolding. Ultimately, through the evaluation of odds ratios based on co-mentions, we measured the strength of association between opioid substances, routes of administration, and drug tampering.\\
\textbf{Results:}
We identified 32 subreddits discussing nonmedical opioid usage from 2014 to 2018 and observed the evolution of interest among over 86,000 Reddit users potentially involved in firsthand opioid usage. We learned the language model of opioid consumption and provided alternative vocabularies for opioid substances, routes of administration, and drug tampering. A data-driven taxonomy of nonmedical routes of administration was proposed. We modeled the temporal evolution of interest in opioid consumption by ranking the popularity of the adoption of opioid substances and routes of administration, observing relevant trends, such as the surge in synthetic opioids like fentanyl and an increasing interest in rectal administration. In addition, we measured the strength of association between drug tampering, routes of administration, and substance consumption, finding evidence of understudied abusive behaviors, like chewing fentanyl patches and dissolving buprenorphine sublingually.\\
\textbf{Conclusions:}
This work investigated some important consumption-related aspects of the opioid epidemic using Reddit data. We believe that our approach may provide a novel perspective for a more comprehensive understanding of nonmedical abuse of opioids substances and inform the prevention, treatment, and control of the public health effects.
\end{abstract}

%%
%% Keywords. The author(s) should pick words that accurately describe
%% the work being presented. Separate the keywords with commas.

%%
%% This command processes the author and affiliation and title
%% information and builds the first part of the formatted document.
\maketitle
%
\input{sections/01-intro.tex}

\input{sections/02-methods}

\input{sections/03-results.tex}

\input{sections/04-discussion.tex}

%\section{Acknowledgments}

%%
%% The next two lines define the bibliography style to be used, and
%% the bibliography file.
\bibliographystyle{ama}
\bibliography{bibliography.bib}

%%
%% If your work has an appendix, this is the place to put it.
\clearpage
\appendix
\input{sections/05-multimedia_appendix.tex}
%\input{sections/09-appendix.tex}
% \clearpage

\end{document}

%% file: sections/01-intro.tex
% !TEX root = ../main.tex

\section{Introduction}\label{sec:introduction}

\subsection{Background}

In the last decade, the United States witnessed an unprecedented growth of deaths due to opioid drugs \cite{cdc_2019}, which sparked from overprescriptions of semisynthetic opioid pain medication such as oxycodone and hydromorphone and evolved in a surge of abuse of illicit opioids like heroin \cite{kolodny2015prescription,compton2016relationship} and powerful synthetic opioids like fentanyl \cite{rose2017prescription,ciccarone2019triple}. Alongside traditional medical, pharmacological, and public health studies on the nonmedical adoption of prescription opioids \cite{mccabe2007motives,agnich2013purple,katz2008internet,butler2008national,butler2011abuse,curtis2019opioid,schifanella2020spatial,richards2020factors,van2015misuse}, several phenomena related to the opioid epidemic have recently been successfully tackled through a digital epidemiology \cite{brownstein2009digital,eysenbach2009infodemiology,salathe2012digital,kim2017scaling} approach. Researchers have used digital and social media data to perform various tasks, including detecting drug abuse \cite{hu2019ensemble,prieto2020detection}, forecasting opioid overdose \cite{ertugrul2019castnet}, studying transition into drug addiction \cite{lu2019redditors}, predicting opioid relapse \cite{yang2018predicting}, and discovering previously unknown treatments for opioid addiction \cite{chancellor2019discovering}. A few recent studies investigated the temporal unfolding of the opioid epidemic in the United States by leveraging complementary data sources different from the official US Centers for Disease Control and Prevention data \cite{kolodny2015prescription,phalen2018fentanyl,zhu2019initial,rosenblum2020rapidly,black2020online} and using social media like Reddit \cite{pandrekar2018social,balsamo2019firsthand}.

Pharmacology research is interested in understanding the consequences of various routes of administration (ROA), that is, the paths by which a substance is taken into the body \cite{mccabe2007motives,kirsh2012characterization,gasior2016routes}, due to the different effects and potential health-related risks tied to them \cite{butler2011abuse,strang1998route,young2010route}. Researchers have estimated the prevalence of routes of administration for nonmedical prescription opioids \cite{butler2008national,kirsh2012characterization,gasior2016routes,katz2011tampering} and opiates \cite{ciccarone2009heroin,carlson2016predictors}; however, they rarely consider less common ROA, such as rectal, transdermal, or subcutaneous administration \cite{gasior2016routes,coon2005rectal}, leaving the mapping of nonmedical and nonconventional administration behaviors greatly unexplored \cite{rivers2017strange,mccaffrey2018natural}. Many of these studies \cite{kirsh2012characterization,gasior2016routes,katz2011tampering} acknowledge that drug tampering, that is, the intentional chemical or physical alteration of medications \cite{mastropietro2014drug}, is an important constituent of drug abuse. The alteration of the pharmacokinetics of opioids through drug-tampering methods, together with unconventional administration, may potentially lead to very different addictive patterns and ultimately have unexpected health-associated risks \cite{strang1998route}. Research has also been focused on developing tamper-resistant and abuse-deterrent drug formulations. However, to the best of our knowledge, no large-scale empirical evidence has been found to unveil the relationships between substance manipulation, unconventional ROA, and nonmedical substance administration.

\subsection{Goals}

This paper seeks to complement current studies widening the understanding of opioid consumption patterns by using Reddit, a social content aggregation website, as the primary data source. This platform is structured into subreddits, user-generated and user-moderated communities dedicated to the discussion of specific topics (Multimedia Appendix, Figure \ref{fig:apx:1reddit}) . Due to fair guarantees of anonymity, no limits on the number of characters in a post, and a large variety of debated topics, this platform is often used to uninhibitedly discuss personal experiences \cite{manikonda2018twitter}. Reddit constitutes a nonintrusive and privileged data source to study a variety of issues \cite{baumgartner2020pushshift,medvedev2018anatomy}, including sensitive topics such as mental health \cite{de2014mental}, weight loss \cite{enes2018reddit}, gender issues \cite{saha2019language}, and substance abuse \cite{lu2019redditors,chancellor2019discovering}. This study’s contributions are manifold. First, leveraging and expanding a recent methodology proposed by Balsamo et al \cite{balsamo2019firsthand}, we identified a large cohort of opioid firsthand users (ie, Reddit users showing explicit interest in firsthand opioid consumption) and characterized their habits of substance use, administration, and drug tampering over a period of 5, years. Second, using word embeddings, we identified and cataloged a large set of terms describing practices of nonmedical opioids consumption. These terms are invaluable to performing exhaustive and at-scale analyses of user-generated content from social media, as they include colloquialisms, slang, and nonmedical terminology that is established on digital platforms and hardly used in the medical literature. We provided a longitudinal perspective on online interest in the opioids discourse and a quantitative characterization of the adoption of different ROA, with a focus on the less-studied yet emerging and relevant practices. We have made available the ROA taxonomy and the corresponding vocabulary to the research community. Third, we quantified the strength of association between ROA and drug-tampering methods to better characterize emerging practices. Finally, we investigated the interplay between the previous 32, dimensions, measuring odds ratios to shed light on the “how” and “what” facets of the opioid consumption phenomenon. We studied a wide spectrum of opioid forms, referred to as “opioids” throughout, ranging from prescription opioids to opiates and illegal opioid formulations. To the best of our knowledge, our contributions are original in both breadth and depth, outlining a detailed picture of nonmedical practices and abusive behaviors of opioid consumption through the lenses of digital data.

%% file: sections/02-methods.tex
% !TEX root = ../main.tex
\section{Methods}

\subsection{Data}

We refer to a publicly available Reddit data set \cite{baumgartner_2020} that contains all the subreddits published on the platform since 2007 \cite{medvedev2018anatomy,Baumgartnerreddit}. In this work, we analyzed the textual part of the submissions and the comments collected from 2014 to 2018. We preprocessed each year separately, filtering out the subreddits with less than 100 comments in a year. We used spaCy \cite{spacy_2020} to remove English stop words, inflectional endings, and tokens with less than 100 yearly appearances. We adopted a bag-of-words model, resulting in a vocabulary of different lemmas for each year. Vocabulary sizes ranged from 300,000 to 700,000 lemmas, with a size growth of approximately 30\%  each year. In Table \ref{tab:dataset_statistics},  the number of unique comments and unique active users per year is reported. A steady growth of approximately 30\%,  per year both in the volume of conversations and in the active user base is observed. \\
All the analyses in this work were performed on a subset of subreddits related to opioid consumption, which were identified using the procedure described here. For space constraints, we restricted the analyses of odds ratios to comments and submissions created during 2018. Similar to a vast body of users’ activities on social media platforms \cite{barabasi2005origin,malmgren2009universality,muchnik2013origins}, the distribution of posts per user shows a heavy tail, with the majority of users posting few comments and the remaining minority (eg, core users and subreddit moderators) producing a large portion of the content. Moreover, a nonnegligible percentage of posts, respectively 25\%,  and 7\%,  of submissions and comments, were produced by authors who deleted their usernames.\\
\begin{centering}
\input{tables/dataset_statistics}
\end{centering}
\subsection{Analytical pipeline}
The methodology adopted in this paper consists of several steps. First, we identified a cohort of opioid firsthand users and the subreddits related to opioid consumption through a semiautomatic algorithm. Second, we trained a word-embedding language model to capture the latent semantic features of the discourse on the nonmedical use of opioids. Third, we exploited the embedded vectors to extend an initial set of medical terms known from the literature, (eg, opioid substance names, ROA, and drug-tampering methods) to nonmedical and colloquial expressions. The terms were organized in a taxonomy that provides a conceptual map on the topic. Moreover, we studied the temporal evolution of the popularity of the main opioid substances and ROA. Ultimately, we measured the strength of the associations between opioid substances, routes of administration, and drug-tampering techniques in 2018.

\subsubsection{Identification of Firsthand Opioid Consumption on Reddit
}\label{sec:sub:retrieval}

We leveraged a semiautomatic information retrieval algorithm developed to identify relevant content related to a topic of interest \cite{balsamo2019firsthand} to collect opioid-related conversations on Reddit yearly. This approach aims at retrieving topic-specific documents by expressing a set of initial keywords of interest; here, it identified relevant subspaces of discussion via an iterative query expansion process, retaining a list of terms $Q_y$ and a list of subreddits $S_y$ ranked by relevance for each year. We merged all the query terms in a set $\Bar{Q} = \bigcup_{y} Q_y $ containing 67 terms. To ensure that the sets $S_y$ were effectively referring to the opioid-related topics and in particular to nonmedical opioid consumption, we performed a manual inspection on the union  of the top 150 subreddits for each year, for a total of 554 subreddits. Three independent annotators, including a domain expert specialized in antidoping analyses, read a random sample of 30, posts, checking for subreddits (1),  mostly focused on discussing the use of opioids, (2) mostly focused on firsthand usage, and (3),  not focused on medical treatments. This yielded a total of 32, selected subreddits, with a Fleiss' $k$ interrater agreement of $k=0.731$, which suggests a substantial agreement, according to Landis and Koch \cite{Landis77}. Multimedia Appendix Table \ref{tab:apx:2subreddits}, presents a complete list of the subreddits broken down by year. Automatic language detection, performed with langdetect \cite{langdetect}, cld2 \cite{pycld2}, and cld3 \cite{pycld3}, showed that the majority of posts (about 90\%) were in English, approximately 5\%,  were non-English messages, and the rest were too short or full of jargon and emojis to algorithmically detect any language. Assuming that an author who writes in one of the selected subreddits is personally interested in the topic, we identified a cohort of 86,445 unique opioid firsthand users involved in direct discussions of opioid usage across the period of study. Summary statistics are reported in Table~\ref{tab:drugs_terms}  In particular, for each year, we computed the number of unique active users and the volume of comments shared, as well as the user’s relative prevalence over the entire amount of Reddit activity. We observed growth from 2014 to 2017, ranging from 15, to 19, users interested in opioid consumption out of every 100,000 Reddit users.
\subsubsection{Vocabulary expansion}\label{sec:sub:embedding}
\begin{figure}[!h]
\centering
  \includegraphics[trim=30 20 5 20, clip,width=.94\linewidth]{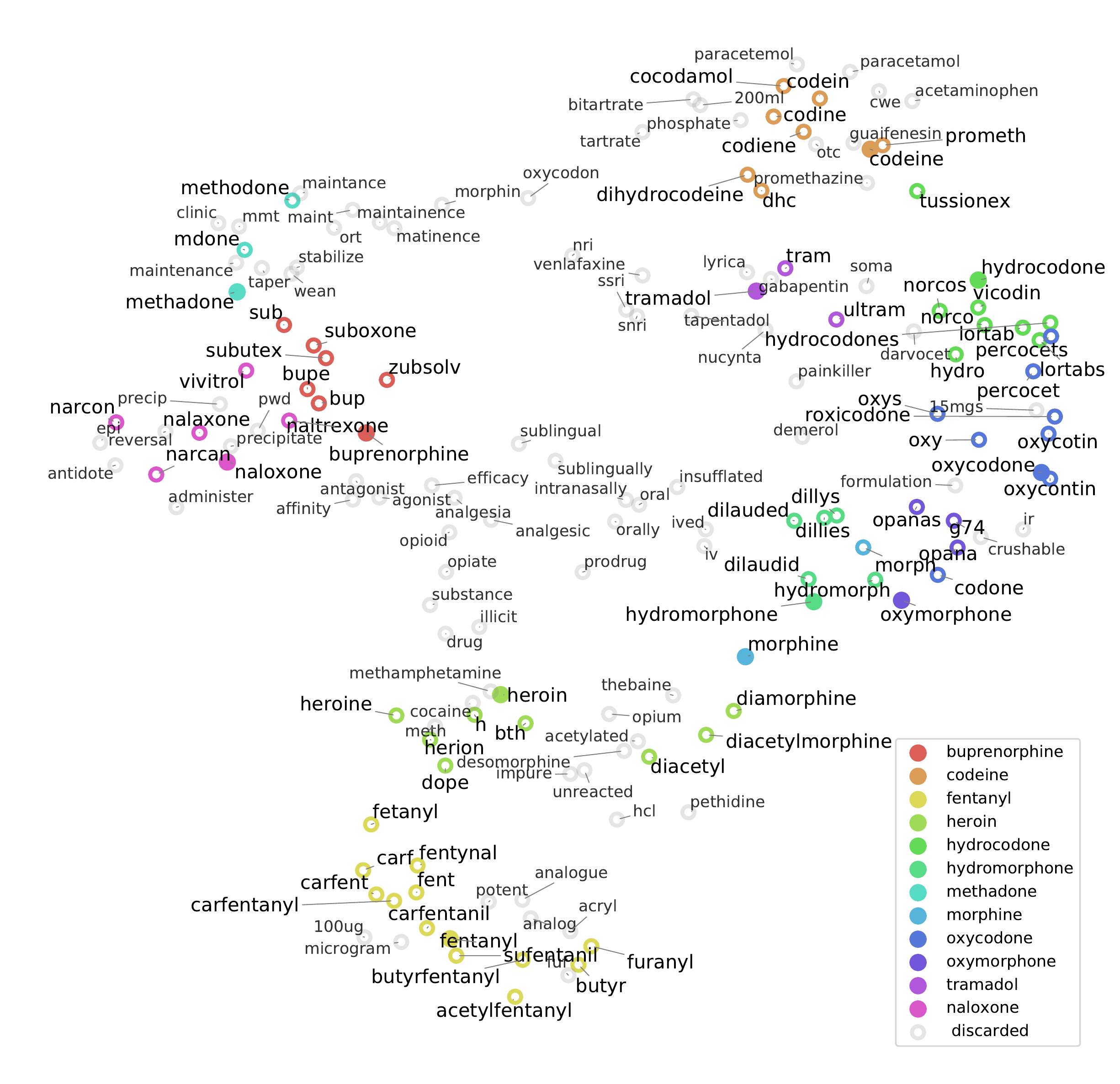}
  \captionof{figure}{Two-dimensional projection of the word2vec embedding, modeling the semantic relationships among terms in the Reddit opioids data set. Filled markers represent the seed terms K. Expansion terms, represented with hollow markers, are colored according to their respective initial term if accepted or in gray if discarded. The nature of the relationships between neighboring terms varies, representing (1) equivalence (eg, synonyms), (2) common practices (eg, the use of methadone for addiction maintenance), or (3) co-use (eg, the cluster of heroin, cocaine, and methamphetamine).}
  \label{fig:embedding}
\end{figure}
The methodology to extend the vocabulary on opioid-related domains with user-generated slang and colloquial forms was implemented in 2, steps. First, we trained a word-embedding model (word2vec \cite{mikolov2013distributed}), which learns semantic relationships in the corpus during training and maps their terms to vectors in a latent vectorial space, with all the comments and submissions in our subreddit data set (relevant training parameters are displayed in Multimedia Appendix Table \ref{tab:apx:3parameters}). Second, starting from a set of seed terms K (eg, a list of known opioid substances), we expanded the vocabulary by navigating the semantic neighborhood $E_w^n=neighbours(w, n)$ of each element  $w \in \Bar{Q} $ in the embedded space, considering the $n=20$, semantically closest elements in terms of cosine similarity. We merged the results in a candidate expansion set, $\Bar{E} = \bigcup_w E_w^n $ , together with the seed terms $K$ if not already included. Based on the knowledge of a domain expert (a clinical and forensic toxicologist) and with the help of search engine queries and a crowdsourced online dictionary for slang words and phrases (Urban Dictionary \cite{urbandictionary}) to understand the most unusual terms, we manually selected and categorized the relevant neighboring terms, obtaining an extended vocabulary $V$. Figure \ref{fig:embedding} shows an example of the expansion procedure in which the high-dimensional vectors are projected to 2 dimensions using the uniform manifold approximation and projection (UMAP) algorithm \cite{mcinnes2018umap-software}. As a sensitivity analysis, we compared the effectiveness of an alternative embedding model (GloVe \cite{pennington2014glove}) for topical coherence. In the case of vocabulary expansion of opioid substance terms, that is, using $K = \Bar{Q} $  as seeds, the 2 models captured 100 terms in common out of their respective candidate terms, with word2vec showing a higher number and a larger percentage of accepted terms (\ref{tab:apx:modelselection}) . Moreover, the volume of comments that included an accepted term was almost double when using the vocabulary of word2vec rather than the vocabulary of GloVe. Hence, we chose word2vec as the reference word-embedding model.

\begin{centering}
\input{tables/embedding_comparison}
\end{centering}
%
%%%%
\subsubsection{Strength of Association Between Opioid Substances, ROA, and Drug Tampering}\label{sec:sub:OR_methods}
%%%%%%%%%%
We evaluated the odds ratios (ORs) to quantify the pairwise strength of the association between substance use and ROA, substance use and drug-tampering methods, and ROA and drug-tampering methods. Under the assumption that co-mention was a proxy for associating a substance to its ROA (or drug-tampering method), we focused on the posts that contained a reference to terms in each domain, evaluating contingency tables and odds ratios. Odds ratios, significance, and confidence intervals were estimated using chi-square tests implemented in the statsmodel Python package \cite{seabold2010statsmodels}, with the significance level set to $\alpha = 0.01$. As a sensitivity analysis, we assessed the effect of the proximity of terms on the characterization of odds ratios. We modified the definition of co-occurrence, introducing a distance threshold $\rho$ at sentence level. We explored the range $\rho \in \{0,...,5\}$, , where $\rho = 0$ indicates that co-occurrence appears within the same sentence, and $\rho > 0$ measures the distance in both directions (eg, $\rho = 1$, for the preceding and consecutive sentences). The value $\rho = \infty$ indicates the scenario in which we considered the entire post as reference. Accordingly, given a threshold $\rho$ in the construction of the contingency table, the co-occurrence event between two terms is conditioned to their distance being less than or equal to $\rho$. Conversely, we considered terms to be separate events in cases of distance above the threshold. It is important to consider that the OR measures do not imply any form of causation but rather surface correlations that could be used in hypothesis formation. To better interpret the results of this analysis, in some cases, manual inspection of the comments mentioning the variables under investigation was performed following the directives on privacy and ethics (see the “Ethics and Privacy” section).
%%%%%%%%%%

%% file: tables/dataset_statistics.tex
\begin{table}[!hpb]
    \centering
    %\small
    \begin{tabular}{lll||ccccc}
    \toprule
    Year &     \makecell{Reddit\\comments, n	} & \makecell{Reddit\\authors, n} &  \makecell{Opiates\\subreddits, n	} &  \makecell{Opiates\\comments, n} &  \makecell{Opiates\\authors, n} &  \makecell{Authors’\\prevalence} \\
    \midrule
    2014  &   545,720,071 &       8,149,234 &                  19 &    386,984 &    12,381 &         0.0015 \\
    2015  &   699,245,245 &      10,673,990 &                  19 &    470,609 &    15,888 &         0.0015 \\
    2016  &   840,575,089 &      12,849,603 &                  25 &    612,619 &    21,791 &         0.0017 \\
    2017  &  1,045,425,499 &      14,219,062 &                  30 &    866,023 &    28,358 &        0.0020 \\
    2018  &  1,307,123,219 &      18,158,464 &                  25 &    919,036 &    33,700 &        0.0019 \\
    \bottomrule
    \end{tabular}
    \caption{Dataset Statistics.}
    \label{tab:dataset_statistics}

\end{table}  

%% file: tables/embedding_comparison.tex
\begin{table}[!ht]
    \centering
    \begin{tabular}{lrrrr}
    \toprule
    {} &   \makecell[ct]{Candidate\\terms, n} &  \makecell[ct]{Accepted\\terms, n (\%)} &   \makecell[ct]{Comments\textsuperscript{a}, \\n}\\
    \midrule
    \textit{word2vec}   &           297 &        128 (43.1) &              225165 \\
    \textit{GloVe} &           369 &        110 (29.8) &             144564 \\
    \bottomrule
    \end{tabular}
    \caption{Comparison of term expansions of opioid substances for the 2 trained models. \textsuperscript{a}Comments in the corpus mentioning at least one term of the respective accepted terms for vocabulary expansion.}
    \label{tab:apx:modelselection}
\end{table}

%% file: sections/03-results.tex
% !TEX root = ../main.tex
\section{Results}

\subsection{Characterizing Interest in Opioids, ROA, and Drug-Tampering Methods}
\begin{figure}[!ht]
\centering
  \includegraphics[width=.9\linewidth]{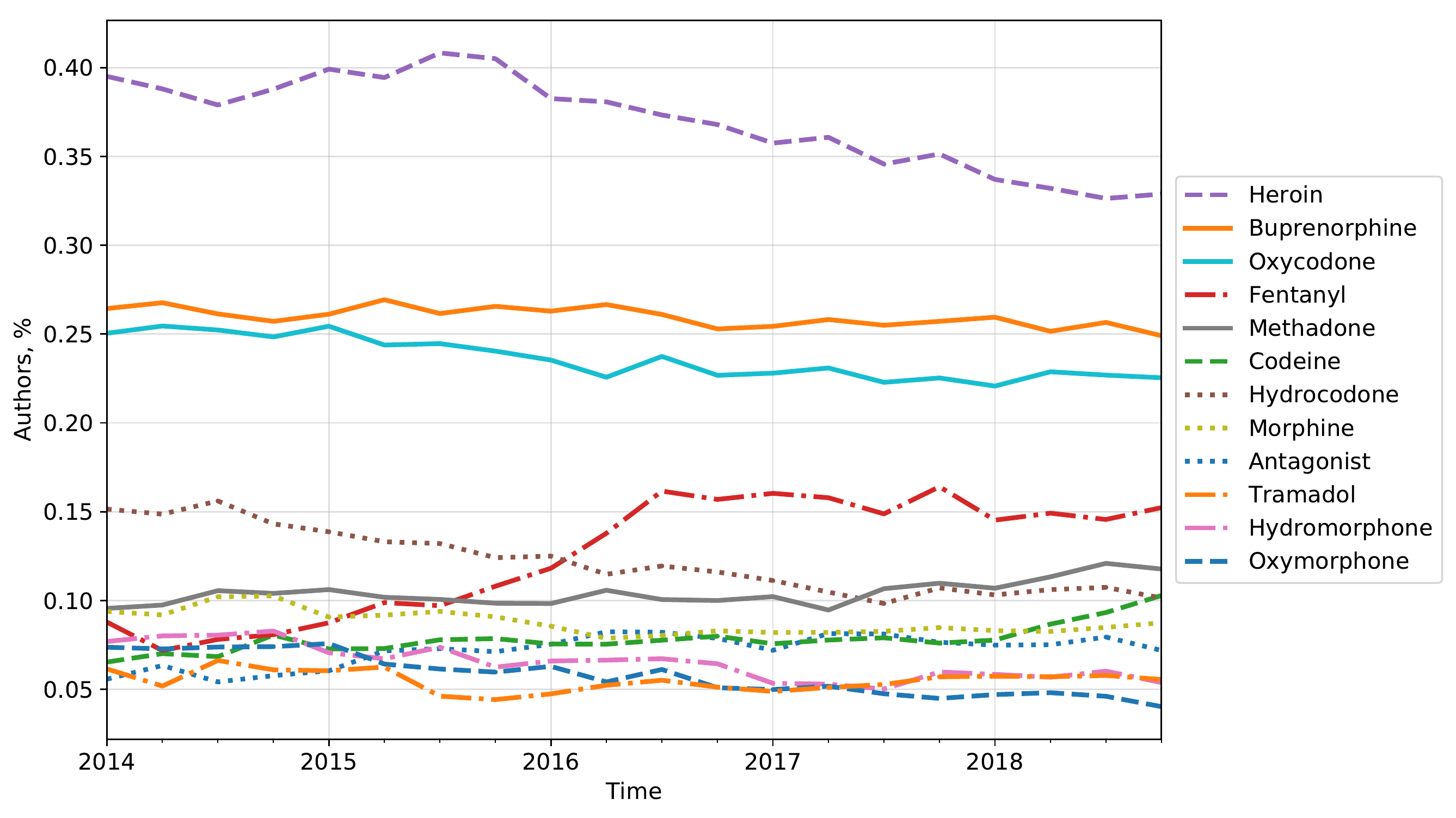}
  \captionof{figure}{Popularity of opioid substances among opioid firsthand users on Reddit. Each line represents the share of opioid firsthand users mentioning an opioid substance, measured quarterly from 2014 to 2018.}
  \label{fig:trend_drugs}
\end{figure}

\begin{figure}
    \centering
    \includegraphics[width =.9\textwidth]{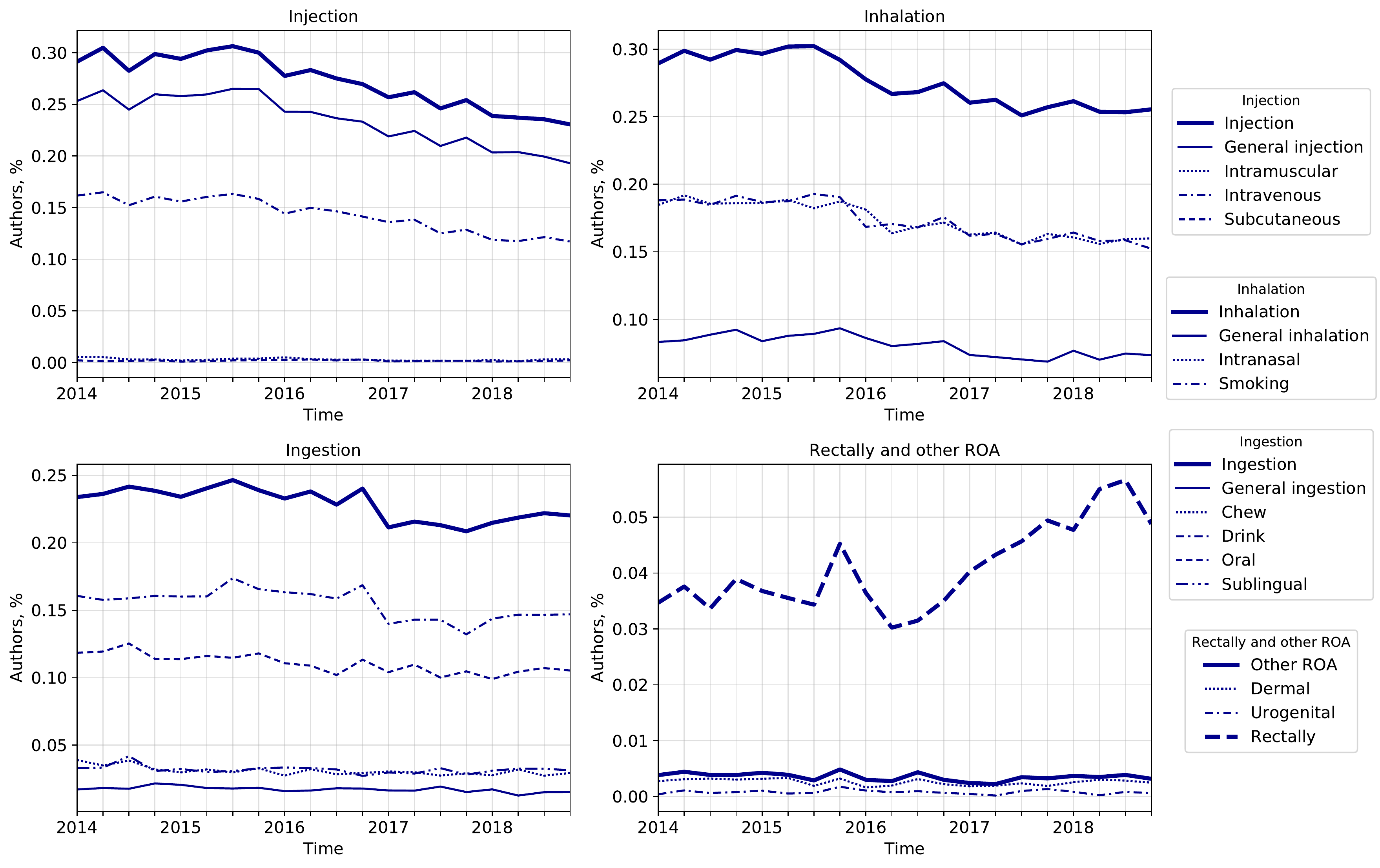}
    \caption{Popularity of routes of administration among opioid firsthand users on Reddit. Each line represents the fraction of opioid firsthand users mentioning an ROA-related term, measured quarterly from 2014 to 2018. Thick lines represent the share of authors mentioning primary ROA, evaluated by aggregating the contributions of all the corresponding secondary ROA. ROA: routes of administration.}  
    \label{fig:trend_roa}
\end{figure}

%%%%%%%%%%
%
\begin{centering}
\input{tables/table_drugs_terms}
\end{centering}
We applied the methodology described in the “Vocabulary Expansion” section to extract and expand domain-specific vocabularies and to characterize the temporal unfolding of interest in different opioid substances, routes of administration, and drug-tampering methodologies. We started from a review of the relevant medical research, collecting an initial set of terms referring to the most common opioid substances, ROA \cite{mccabe2007motives,butler2011abuse,kirsh2012characterization,young2010route,coon2005rectal,rivers2017strange,mastropietro2014drug,hart2014me,surratt2017heroin}, and drug-tampering methods \cite{mastropietro2014drug,hart2014me}. We expanded the original set with neighboring terms in a low-dimensional embedding space, and the outputs were reviewed and organized by a domain expert. The resulting vocabulary for opioid substances is shown in Table \ref{tab:drugs_terms}.  It is worth noting that the vocabulary expansion procedure considerably increased the richness of the terminology related to the domain of interest and, consequently, the volume of conversations on Reddit that contained these terms. For example, for the heroin category, we observed a 62\%,  growth in the retrieved relevant conversations (Table \ref{tab:drugs_terms}). We investigated the temporal unfolding of the popularity of the opioid substances, measured as the fraction of authors mentioning a substance over the entire opioid firsthand user base, for each trimester from 2014 to 2018. A binary characterization of the mentioning behavior at the user level was considered to discount potential biases due to users with high activity. We also provided a relative measure of popularity to account for the constantly increasing volume of active users on Reddit. Figure \ref{fig:trend_drugs} shows a decrease in the usage of heroin and a rise in fentanyl and codeine.
The resulting vocabulary for routes of administration was further organized in a 2-level hierarchical structure, reported in Table \ref{tab:roa_terms}.  It is worth noting that the taxonomy does not have a strict medical interpretation, nor was it intended to be a comprehensive review. However, it can give structure to otherwise unstructured collections of words and help in the interpretation of the results.
Figure \ref{fig:trend_roa} shows the estimated temporal evolution of the relative popularity of the routes of administration from 2014 to 2018, measured in quarterly snapshots. Finally, we extracted and organized the vocabulary related to drug-tampering techniques, as shown in Table \ref{tab:trans_terms}.  In this paper, we considered the act of chewing pills a second-level route of administration under the ingestion category \cite{katz2008internet,kirsh2012characterization,gasior2016routes} instead of a drug-tampering method, as some research might suggest \cite{mastropietro2014drug}.
%%%%%%%%%%%
%

\begin{centering}
\input{tables/roa_taxonomy}
\end{centering}
\begin{centering}
\input{tables/tampering_terms}
\end{centering}
\subsection{Characterizing the Associations Between Opioid Substances, ROA, and Drug Tampering}\label{sec:sub:aor_roa_drugs}
%%%%%
\begin{figure}[b]
\centering
{\includegraphics[width=.95\textwidth]{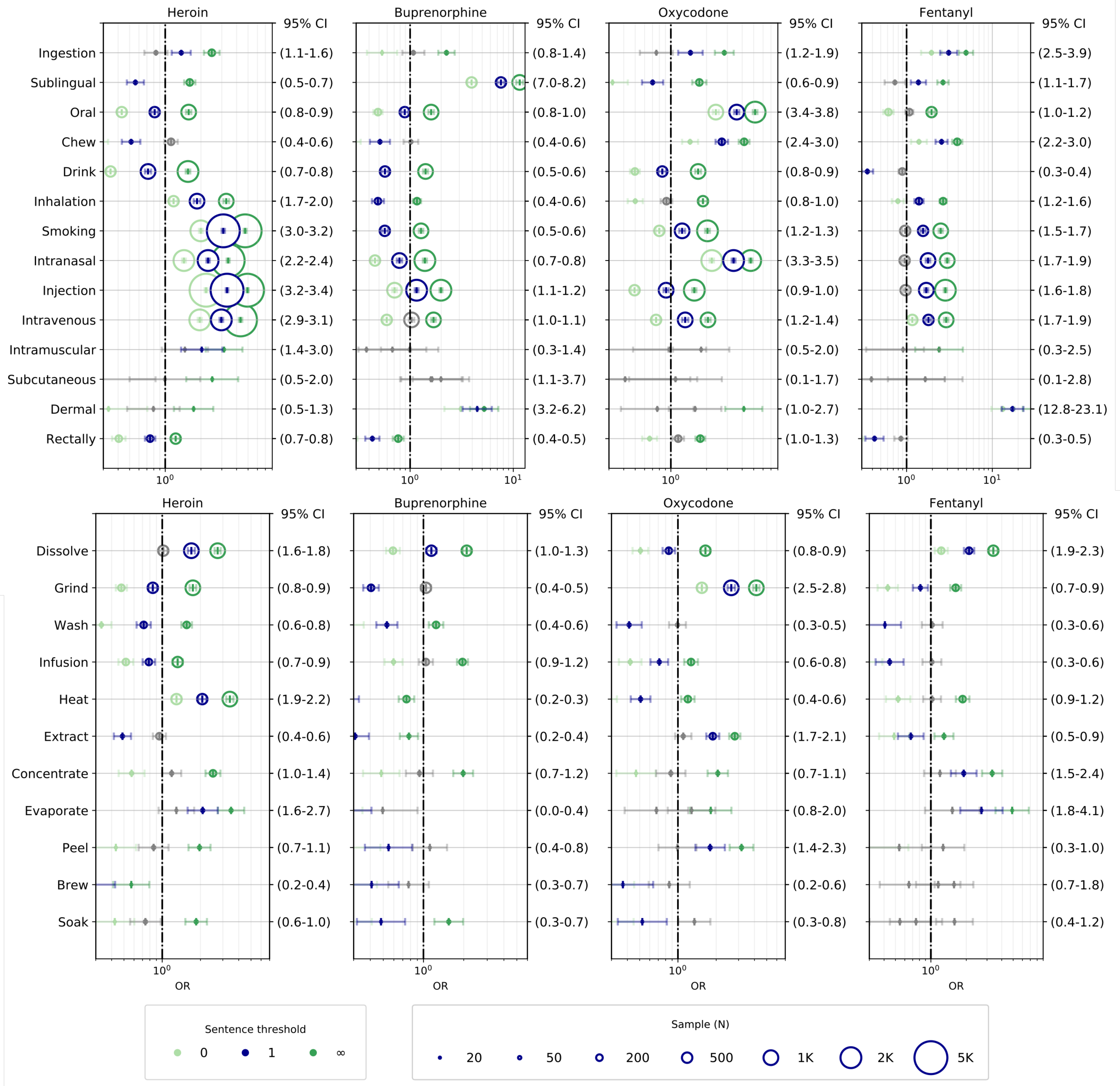}
\phantomsubcaption\label{fig:sub:OR_drugs_trans}}
\caption{Odds ratios of the most widespread opioid substances with routes of administration (top row) and drug-tampering methods (bottom row). Labels on the right axis report the confidence interval at $\rho =1$. OR: odds ratio.}
\label{fig:OR_drugs_roa_trans}
\end{figure}
\begin{figure}[!htbp]
    \centering
    \includegraphics[width=.95\textwidth]{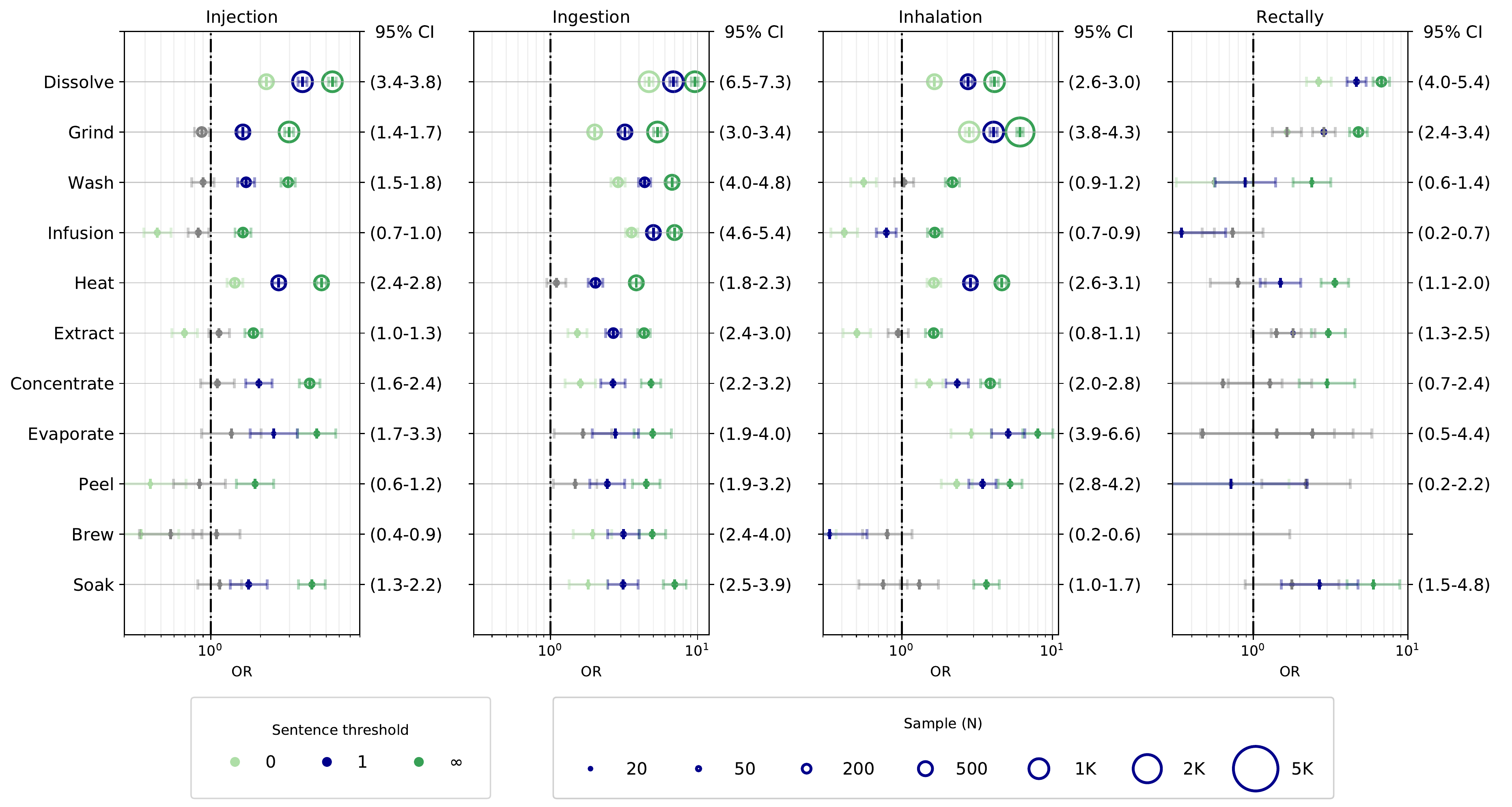}
    \caption{Odds ratios of the primary routes of administration (excluding other routes of administration) and drug-tampering methods. Labels on the right axis report the confidence interval at $\rho =1$. OR: odds ratio.}
    \label{fig:OR_trans_roa}
\end{figure}

%
%%%%%%%%%%
To investigate the strength of association between routes of administration, drug tampering, and opioid substances and to shed light on the interplay between the “how” and the “what” dimensions of opioid consumption, we estimated the ORs, 95\% confidence intervals, P values, and volume of the co-mentions among substances, routes of administration, and drug-tampering methods. The number of sentences in Reddit posts vary greatly, but the posts are generally quite short (approximately 50\% of them have 2 sentences or less, as seen in Multimedia Appendix Figure \ref{fig:apx:4cum}). However, as about 20\% of posts have more than 10 sentences, one should be cautious in adopting a bag-of-words approach to measure co-occurring terms. To limit the chance of including spurious correlations due to the co-occurrences of terms far apart in the posts, we conservatively selected $\rho = 1$, (ie, considering only the co-occurrence of terms within a sentence or in the first adjacent sentences) for computing the OR. Figure \ref{fig:OR_drugs_roa_trans} shows in blue the results of the analysis at $\rho = 1$,  matchin 4 of the main widespread substances (ie, heroin, buprenorphine, oxycodone, and fentanyl) with the secondary ROA (upper panel) and the drug-tampering techniques (lower panel). Figure \ref{fig:OR_trans_roa} shows the odds ratios of primary ROA and drug-tampering methods. For reference, the green markers represent the ORs obtained at $\rho = 0$ and $\rho = \infty$ for the same categories. Multimedia Appendix Figures \ref{fig:apx:5ordrugsroa},\ref{fig:apx:6OR_drugs_trans_all},\ref{fig:apx:7OR_trans_roa_all}, provide the complete set of results for all the substances identified and the secondary ROA. Due to the low representativeness of intrathecal and urogenital ROA with most of the tampering-related terms, we omitted those categories from the analysis. In the plots, the associations that are not statistically significant (P>.01) are reported in gray, and the horizontal lines indicate the OR and the 95\% confidence interval. The radius of the circle is proportional to the sample of co-mentions and the dashed vertical line corresponds to an OR of 1, for reference.
%%%%%%%%%%
%

%% file: tables/table_drugs_terms.tex
\begin{table}[hb]
    \small%footnotesize	
    \centering
    %. $\frac{|\Bar{E_{s}}|-|\Bar{Q_{s}}|}{|\Bar{Q_{s}}|\cup|\Bar{E_{s}}|}$
    \begin{tabular}{llcc}
    \toprule
    Substance & Terms   & $\Delta{Volume, \%}$\\
    \midrule
    Heroin & \makecell[lt]{ \textbf{bth}, diacetylmorphine, diamorphine,\\ dope, \textbf{ecp}, goofball, goofballs, \\gunpowder, h, \textbf{herion}, \textbf{heroin}, heroine, heron, smack,\\ speedball, speedballing, \textbf{speedballs}, tar } &            62 \\
    Buprenorphine & \makecell[lt]{ bup, \textbf{bupe}, \textbf{buprenorphine}, \textbf{butrans}, sub, \textbf{suboxone}, \\\textbf{subutex}, zub, \textbf{zubsolv} } &            61 \\
    Hydrocodone & \makecell[lt]{ hydro, \textbf{hydrocodone}, \textbf{hydrocodones}, \textbf{lortab},\\ \textbf{lortabs}, \textbf{norco}, \textbf{norcos}, tuss, \textbf{tussionex}, vic,\\ vicoden, \textbf{vicodin}, \textbf{vicodins}, \textbf{vicoprofen}, \textbf{vics},\\ vikes, viks, \textbf{zohydro} } &            38 \\
    Codeine & \makecell[lt]{ cocodamol, \textbf{codein}, \textbf{codeine}, \textbf{codiene}, codine, \\dhc, \textbf{dihydrocodeine}, prometh, sizzurp, syrup } &            28 \\
    Oxymorphone & \makecell[lt]{ g74, \textbf{opana}, opanas, \textbf{oxymorphone}, panda } & 25 \\
    Tramadol & \makecell[lt]{ desmethyltramadol, dsmt, tram, \textbf{tramadol},\textbf{ultram}} &            22 \\
    Hydromorphone & \makecell[lt]{ dil, dilauded, \textbf{dilaudid}, dilaudids, \textbf{dillies}, dilly, dillys, \\\textbf{diluadid}, \textbf{hydromorph}, \textbf{hydromorphone} } &            21 \\
    Oxycodone & \makecell[lt]{ 15s, 30s, codone, contin, ms, oc, ocs, \textbf{oxy}, \textbf{oxycodone}, \\\textbf{oxycontin}, oxycontins, \textbf{oxycotin}, \textbf{oxys},\\ \textbf{perc}, \textbf{percocet}, \textbf{percocets}, percoset, percosets,\\ \textbf{percs}, perk, \textbf{roxi}, \textbf{roxicodone}, \textbf{roxie}, \textbf{roxies},\\ \textbf{roxis}, \textbf{roxy}, \textbf{roxycodone}, \textbf{roxys} } &            14 \\
    Morphine & \makecell[lt]{ kadian, morph, \textbf{morphine} } &            5 \\
    Fentanyl & \makecell[lt]{ \textbf{acetylfentanyl}, butyr, butyrfentanyl, carf,  carfent, \\\textbf{carfentanil}, carfentanyl,\textbf{duragesic}, \textbf{fent}, \textbf{fentanyl}, fents, \\fentynal, fetanyl, furanyl, sufentanil, u47700 } &            4 \\
    Antagonist & \makecell[lt]{ \textbf{nalaxone}, \textbf{naloxone}, naltrexone, \textbf{narcan},  narcon, revia,\\ viv, \textbf{vivitrol} }  &            1 \\
    Methadone & \makecell[lt]{ mdone, \textbf{methadone}, \textbf{methodone} } &            1 \\
   
    \bottomrule
    \end{tabular}  
    \caption{Vocabulary of opioid substances. Starting from a candidate expansion set $\Bar{E}$, comprising 297 unique terms, the final expansion terms considered equivalent to a substance were gathered in the same class. Terms in $\Bar{Q}$ are highlighted in bold. The increase in the volume of occurrences of a substance using the terms in the expanded vocabulary compared with only using the terms in $\Bar{Q}$.} 
    \label{tab:drugs_terms}
\end{table}

%% file: tables/roa_taxonomy.tex
\begin{table}[htb]
    \small%footnotesize	
    \centering
    \begin{tabular}{lll}
    \toprule
    Primary ROA & Secondary ROA                &                                Terms \\
    \midrule
    Ingestion & Oral & \makecell[lt]{bolus, buccal, gulp, mouth, mouthful, \textbf{oral},\\ orally, \textbf{swallow}} \\
     & Sublingual & \textbf{sublingual}, sublingually, tongue, tounge \\
     & Drink & chug, drink, pour, pourin, \textbf{sip}, sipper, sippin, swig, swish \\
     & Chew & \textbf{chew}, chewy, chomp, gum \\
     & General Ingestion & \textbf{ingest}, ingestion\\
    
    Inhalation & Intranasal& \makecell[lt]{intranasal, intranasally, nasal, nasally, nose, nostril,\\ rail, \textbf{sniff}, sniffer, sniffin, snoot, snooter, \textbf{snort}, snorter,\\ tooter} \\
    
     & General Inhalation& \makecell[lt]{breath, breathe,dab, exhale, inhalation, \textbf{inhale}, insufflate,\\ insufflated, insufflating, insufflation, puff, toke, tokes, vap,\\ vape, vaped, vapes, vaping, vapor, vaporise, vaporize,\\ vaporizer, vapour }\\
    
     & Smoking& bong, fume, hookah, pipe, \textbf{smoke}, smoker, smokin, spliff \\
    
    Injection & Intramuscular& deltoid, imed, iming, \textbf{intramuscular}, intramuscularly \\
    
     & Subcutaneous& \textbf{subcutaneous}, subcutaneously, subq \\
    
     & Intravenous& \makecell[lt]{arterial, bloodstream, \textbf{intravenous}, intravenously,\\ \textbf{iv}, ivd, ived, iving, ivs, vein, venous} \\
    
     & General Injection & \makecell[lt]{bang, \textbf{inject}, injectable, injection, \\ parenteral, shoot, shot }\\
    
    Rectally & Rectally & \makecell[lt]{anal, anally, \textbf{boof}, boofed, boofing, bunghole, butt,\\ pooper, \textbf{rectal}, rectally }\\
    
    Other ROA & Dermal & cutaneous, dermis, \textbf{transdermal}, transdermally \\
    
     & Urogenital & vaginal \\ 
    
     & Intrathecal & intrathecal \\
    
    \bottomrule
    \end{tabular}
    \caption{Taxonomy defining the ROA categories and their corresponding terms. Primary ROA include all the expansion terms considered for the appropriate secondary ROA (original candidate expansion set comprised 199 unique terms). Seeds in $K$ are highlighted in bold.}
    \label{tab:roa_terms}

\end{table}

%% file: tables/tampering_terms.tex
\begin{table}[htbp]
    \small%footnotesize	
    \centering
    \begin{tabular}{ll}
    \toprule
    Transformation &               Terms \\
    \midrule
    
    Brew & \textbf{brew}, brewer, \textbf{homebrew} \\
    
    Concentrate & \textbf{concentrate}, concentrate,concentration, purify \\
    
    Dissolve & \makecell[lt]{ desolve, dilute, disolve, disolved, disolves, \textbf{dissolve},\\ solute, solution, soluble, soluable,  }\\
    
    Evaporate & evap, evaporate \\
    
    Extract &  \textbf{cwe}, \textbf{extract}, extraction \\
    
    Grind & \makecell[lt]{ chop, \textbf{crush}, crushable, crusher, \textbf{grind}, grinded, grinder,\\ ground, pulverize }\\
    
    Heat & boil, \textbf{heat}, melt, microwave, overheat, simmer \\
    
    Infusion & infuse, \textbf{infusion}, tea, tincture \\
    
    Peel & peal, peel, shave\\
    
    Soak & \textbf{soak}, submerge \\
    
    Wash & rewash, rinse, \textbf{wash} \\
    \bottomrule
    \end{tabular}
    \caption{Vocabulary of drug-tampering methods. Expansion terms referring to the same drug-tampering method are grouped in the corresponding transformation classes (original candidate expansion set comprised 179 unique terms). Seed terms $K$ are highlighted in bold.}
    \label{tab:trans_terms}
\end{table}

%% file: sections/04-discussion.tex
\section{Discussion}
\subsection{Opioid Interest on Reddit}
In this work, we identified over 3 million comments on 32 subreddits focused on discussing practices and implications of firsthand opioid use. We also selected a cohort of over 86,000 Reddit users interested in this topic. Such a large data set allowed us to assess the magnitude of the online interest in opioids and model its evolution during the 5 years of study, sadly verifying its rapidly increasing popularity. By the end of 2018, the opioid epidemic remained an escalating public health threat, and at the time of writing, the opioid crisis is still calling for countermeasures at scale. Hence, we believe our large data set may constitute a valid alternative source to advise decision making and a valuable starting point for future infodemiology research.

\subsection{Vocabulary expansion}
By observing the vocabularies in Tables \ref{tab:drugs_terms},\ref{tab:roa_terms},\ref{tab:trans_terms} resulting from the expansion algorithm, we can ascertain the importance of enriching domain expertise with user-generated content and observe that some common features are captured across categories. Our method was able to detect synonyms and common short names, very specific acronyms (eg, “cwe” for cold water extraction \cite{bausch2012tamper}), slang expressions like “sippin” (often used when referring to the act of drinking codeine mixtures \cite{hart2014me}), nicknames (eg, “panda” for oxymorphone), and polypharmacy instances (eg, “speedball” and “goofball” \cite{ellis2018twin}). The vocabulary expansion underlines the use of prescription dosages (usually stamped on the tablets) in place of the commercial names of the substances (eg, "30s” for oxycodone). Moreover, we deduced that opioid firsthand users discussed variants of the substances (eg, “bth” and “ecp” for black tar heroin and East Coast powder), research chemical equivalents (eg, “u47700” \cite{prekupec2017misuse}), and formulations intended for veterinary use (eg, sufentanil, carfentanil). ROA vocabulary included and categorized both medical terms, adding terms scarcely considered in previous studies, like “vaping,” and nonmedical or unconventional administration terms, such as “chewing,” “snorting,” “smoking,” and “boofing” \cite{rivers2017strange}. Our taxonomy also enabled us to disambiguate common primary ROA, such as injection and ingestion, into specific secondary ones, like subcutaneous \cite{rivers2017strange} and sublingual administrations. Finally, the drug-tampering vocabulary captured tampering methods that modify the physical status of the substances, like crushing and peeling, and some methods aiming at altering the chemical characteristics of the substances, like dissolving, washing, and heating \cite{mastropietro2014drug}. We believe that even if this vocabulary might not be exhaustive of all drug-tampering methods, it offers a novel evidence-based perspective on the topic compared with the existing literature. The expanded vocabularies proved essential to fully incorporating the language complexity of online discussions and taboo behaviors \cite{allan2006forbidden} into at-scale analyses. Hopefully, our contribution might be useful in the future to find and understand new abusive behaviors that are discussed online, ultimately driving future research to yield more effective prevention methods.
%%%%%%%%%
\subsection{Adoption Popularity of Opioid Substances and ROA}
Considering the share of users mentioning a term to be a proxy of firsthand involvement in opioid-related activities and including topic-specific terminology, the longitudinal views in Figures \ref{fig:trend_drugs} and \ref{fig:trend_roa} can be used to rank the popularity of nonmedical usage of opioid substances and ROA and their adoption trends. Ranking the substances by average share, we can see that heroin is by far the most popular substance, mentioned on average by 1, in every 3 users. Its share of users, though, is steadily decreasing, with a loss of 10\%  reported in state-specific findings by Rosenblum et al \cite{rosenblum2020rapidly}. Buprenorphine and oxycodone were the most mentioned prescription opioids; they showed fairly static behavior, while hydrocodone importance decreased over time \cite{black2020online}, possibly due to more stringent prescription regulation starting in 2014 \cite{deaHydrocodone}. Fentanyl showed the most abrupt behavior, dramatically increasing since 2016. Its volume of mentions in 2018 increased by almost 1.5 times compared with 2014, confirming it as one of the most recent threats \cite{ciccarone2019triple,black2020online}. In contrast, we did not find evidence of drastic changes in oxymorphone adoption after its partial ban in 2017 \cite{FDAoxymorphone}. ROA adoption was led by injection and inhalation, which were the most popular ROA across the years, mentioned by 1 of every 3 authors at their peak. These were followed closely by ingestion. Rectal use and other ROA involved, on average, a significantly lower share of users,  around 5\% and less than 1\%, respectively. Nevertheless, rectal administration has shown a sharp increase in popularity since 2016, almost doubling its share. Administration through inhalation was equally staggered by the intranasal and smoking categories of secondary ROA, strong indicators that this route of administration is indeed capturing nonmedical use of opioids. This work on understanding which substances are currently gaining popularity may give prevention programs a strategic advantage, especially if consumption trends can be localized geographically \cite{schifanella2020spatial,balsamo2019firsthand,basak2019detection}, focusing the interventions needed to prevent early adoption of emerging dangerous substances like fentanyl. Moreover, tracking the evolution of interest in prescription opioids might be useful for evaluating the efficacy of ban policies, as in the case of oxymorphone. Understanding which ROA are the most adopted might eventually help address targeted campaigns informing users on safer practices, develop better tamper-resistant prescription drugs, and ultimately better inform the health system of the health risks specific to opioid adoption.
%%%%%%%%%%%%%
\subsection{Characterizing the Association Between Substance Consumption, ROA, and Drug-Tampering Methods}
%%%%%%%%%%%
By jointly considering the results of the odds ratios in Figures \ref{fig:OR_drugs_roa_trans} and \ref{fig:OR_trans_roa} and Multimedia Appendix Figures \ref{fig:apx:5ordrugsroa},\ref{fig:apx:6OR_drugs_trans_all},\ref{fig:apx:7OR_trans_roa_all}, we can outline complex preferences for the nonmedical use of opioids, triangulating substance use, ROA, and drug-tampering methods. We noticed that the majority of substances exhibited more than one high odds ratio, both with ROA and drug-tampering methods, meaning that such substances might be consumed by users in multiple nonexclusive ways. Our analysis shows that for the most part, the expected medical and nonmedical routes of administration of each substance (ie, intended ROA or known abusive administration) had high odds ratios. For prescription opioids, oral (medical) use was often confirmed (eg, oxycodone: OR 3.6, 95\% CI 3.4-3.8), while intranasal administration was often the preferred nonmedical ROA, followed by injection, especially intravenous administration (eg, hydromorphone: OR 9.1, 95\% CI 8.6-9.8) \cite{gasior2016routes,omidian2015routes}. As expected, heroin appeared to be most likely consumed through injection (OR 3.3, 95\% CI 3.2-3.4) or smoking, if heated up on aluminum foil (OR 3.1, 95\% CI 3.0-3.2). Heroin was the only substance that showed high correlations with this administration route. It was also reported to be snorted \cite{surratt2017heroin}.\\
Besides confirming and quantifying some known behaviors, our analysis can provide additional insights on the nonmedical use of intended routes of administration. In accordance with the literature \cite{kirsh2012characterization,gasior2016routes,mccaffrey2018natural,butler2013abuse}, we found evidence that abuse of prescription opioids may be associated with chewing the pills (eg, oxycodone: OR 2.7, 95\% CI 2.4-3.0). From the analysis of ROA and drug-tampering methods, it appears that nonmedical oral administration was correlated with dissolving (OR 9.7, 95\% CI 9.0-10.4), grinding, and washing the substances. In some cases, oral and chewing-related misuse of prescription opioids simply consisted of peeling (OR 5.1, 95\% CI 2.6-9.9) the external coating, which is usually hard to chew or responsible for the extended-release effect. Even though some formulations, such as Opana ER (oxymorphone hydrochloride extended-release tablets; Endo Pharmaceuticals), are known to be tamper resistant to crushing, users can peel the tablets to get rid of the extended release coating for higher recreational effects. Injection usually requires that the substance be dissolved (OR 3.5, 95\% CI 3.2-3.7), while inhalation requires that the substance be ground to powder, especially for intranasal abuse (OR 6.7, 95\% CI 6.3-7.1).\\
Our method ultimately found evidence of unconventional nonmedical administration for most of the substances. We found a high correlation between dissolving and intranasal administration (OR 4.1, 95\% CI 3.8-4.4), which may indicate the adoption of “monkey water,” the practice of dissolving soluble substances, like tar heroin and fentanyl patches, and using the resulting liquid as a nasal spray \cite{ciccarone2009heroin}. Fentanyl patches were also consumed in other unforeseen ways; an unexpectedly high OR of fentanyl and chewing (OR 2.6, 95\% CI 2.2-3.0) suggests that prescription patches intended for transdermal use may be chewed for nonmedical use. Our analyses revealed the high odds of abuse of codeine via drinking (OR 4.0, 95\% CI 3.7-4.3) codeine syrup, made by extracting or brewing the cough suppressants (OR 14.1, 95\% CI 11.5-17.2) and forming the so-called lean or purple drank \cite{agnich2013purple,hart2014me,cherian2018representations}.\\
Buprenorphine, usually administered sublingually in its formulations without an antagonist, such as Subutex (buprenorphine; Indivior), and orally in combination with naloxone in the form of pills, such as Suboxone (buprenorphine-naloxone; Indivior) and Zubsolv (buprenorphine-naloxone; Orexo), measured exceptionally high odds of sublingual administration (OR 7.6, 95\% CI 7.0-8.2). Evidence of nonmedical use of buprenorphine was also found in the association between dissolving and sublingual use (OR 18.9, 95\% CI 16.8-21.3). Opioid firsthand users know that the opioid antagonist in buprenorphine-naloxone compounds has low bioavailability if dissolved under the tongue; hence, to achieve higher opioid effects and eliminate the antagonist, these compounds are generally taken sublingually and not through other ROA, with which buprenorphine shows negative associations. Finally, our study shows that rectal administration is a viable and unforeseen option for the nonmedical use of some opioids, resulting in higher recreational effects, especially with hydromorphone (OR 5.2, 95\% CI 4.6-6.0), morphine, and oxymorphone. Rectal administration showed high correlations with the dissolving, grinding, and soaking drug-tampering methods, possible indicators of an unconventional route of administration, largely overlooked, which involves dissolving the substances in liquid water or alcohol (ie, “butt-chugging”) \cite{rivers2017strange,el2015butt}. Subcutaneous administration was only weakly associated with morphine, suggesting that the practice of “skin popping” \cite{coon2005rectal}, which consists of injecting the substance in the tissues under the skin, is potentially not widespread.\\
The complex interactions between substance use, routes of administration, and drug tampering that can be unveiled with our methodology provide a broad yet detailed perspective on the nonmedical use of opioids, evidencing abusive behaviors in which unconventional ROA and drug tampering play a key role. Knowledge about abusive behaviors could be taken into consideration by physicians during treatment programs, allowing them to favor opioid medications that are less likely to be transformed and abused. Our results should be addressed with effective health policies, driving future clinical research to better focus its efforts on understanding health-related risks and guiding the production of new tamper-resistant and abuse-deterrent opioid formulations.
%%%%%%%%%%
%
\subsection{Limitations and future work}
We acknowledge some limitations in the present research. The population sampled on Reddit might have intrinsic social media biases, and it is likely not representative of the general population (eg, for gender, age, or ethnicity). Moreover, since we enrolled the users in our cohort based on their engagement in subcommunities focusing on firsthand use of opioids, we cannot exclude the possibility that in some cases, such users might have been reporting secondhand experiences, disseminating general news, or discussing intended medical drug use for pain management. We must also consider that the selected individuals were not clinically diagnosed with opioid use disorder. Future work will be devoted to building a classifier at the user level to identify individuals with opioid use disorder. We are aware that Reddit data have some gaps~\cite{gaffney2018caveat}, but since the incompleteness mostly affects the years before 2010, we consider the overall results of our work to not be significantly biased. Other limitations are related to the analytic pipeline, where we narrowed our text analysis to term counts and co-occurrences, which might have produced spillover effects in comments discussing multiple topics and could have amplified the strength of cross-associations. Future work will include n-grams and more context-based language models. Finally, it is worth stressing that the measure of association through odds ratios should not be intended by any means as an indication of causal effects. This work is an observational study focusing on the characterization of a complex and faceted social phenomenon rather than the identification of determinants or interventions, and it shares the strengths and limitations of correlational studies, especially in medical research.
\subsection{Ethics and Privacy}\label{sec:sub:ethics}
Given the sensitive nature of the information shared, including users’ vulnerabilities and personal information, privacy and ethical considerations are paramount. In this work, we followed the guidelines and directives in Eysenbach and Till \cite{eysenbach2001ethical}, which describe recommendations to ethically conduct medical research with user-generated online data, and we relied on the vast experience of research works dealing with sensitive data gathered on social media \cite{saha2019language,moreno2013ethics,chancellor2019taxonomy,ramirez2020detection,hswen2018monitoring}. The researchers had no interactions with the users and have no interest in harming any, and the analyses were performed and reported in the spirit of knowledge, prevention, and harm reduction. In this direction, it is worth noting that the subreddits under study are of public domain, are not password protected, and have thousands of active subscribers; users were fully aware of the public nature of the content they posted and of its free accessibility on the web. Moreover, Reddit offers pseudonymous accounts and strong privacy protection, making it it unlikely that the true identity of a user can be recovered. Nevertheless, in order to further protect the privacy and anonymity of the users in our data set, all information about the names of the authors was anonymized before using the data for analysis. Moreover, every analysis performed was intended to provide aggregated estimates aimed at research purposes, and this work did not include any quotes or information that focused on single authors. Following the directives in Eysenbach and Till \cite{eysenbach2001ethical}, our research did not require informed consent.

\section{Conclusions}
In this work, we characterized opioid-related discussions on Reddit over 5 years, involving more than 86,000 unique users, and focused on firsthand experiences and nonmedical use. To address the complexity of the language in social media communications, especially in the presence of taboo behaviors such as drug abuse, we gathered a large set of colloquial and nonmedical terms that covered most opioid substances, routes of administration, and drug-tampering methods. We were able to characterize the temporal evolution of the discourse and identify notable trends, such as the surge in the popularity of fentanyl and the decrease in the relative interest in heroin. Focusing on routes of administration, we extended pharmacological and medical research with an in-depth characterization of how opioids substances are administered, since different practices imply different effects and potential health-related risks. We proposed a 2-layer taxonomy and corresponding vocabulary that enabled us to study both medical and recreational routes of administration. We demonstrated the presence of conventional nonmedical ROA (eg, intranasal administration and intravenous injection) and the spread of less conventional practices (eg, an increasing trend in rectal use). In particular, with reference to nonconventional ROA, we characterized for the first time at scale the phenomenon of drug tampering, which could have an impact on health outcomes, since it alters the pharmacokinetics of medications. The interplay between these dimensions was systematically characterized by quantitatively measuring the odds ratios, providing an insightful picture of the complex phenomenon of opioid consumption as discussed on Reddit.
\section*{Acknowledgments}
PB acknowledges support from the Intesa Sanpaolo Innovation Center. The funder had no role in the study design, data collection, analysis, decision to publish, or preparation of the manuscript. RS was partially supported by the project Countering Online Hate Speech Through Effective On-line Monitoring, funded by Compagnia di San Paolo.

%% file: sections/05-multimedia_appendix.tex
\section*{Multimedia Appendix}

\begin{figure}[!htbp]
    \centering
    \includegraphics[width=.8\textwidth]{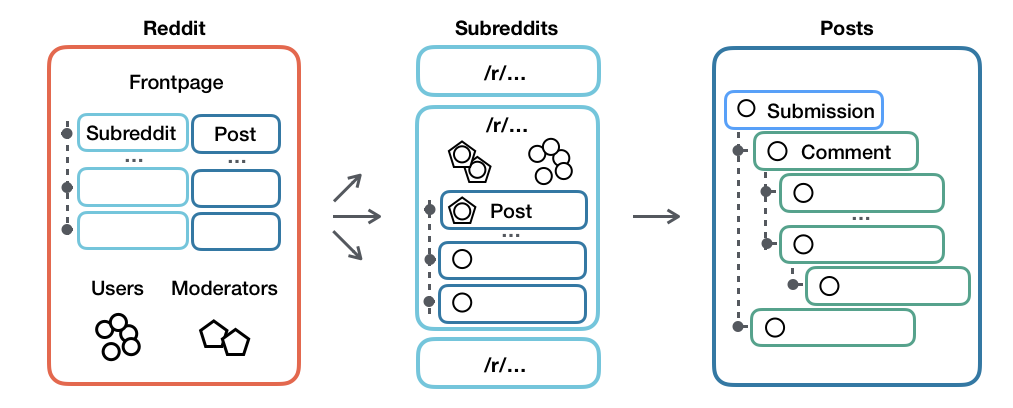}
    \caption{Schematic representation of the structure of Reddit. Reddit's most common access point is the front page, where the most relevant content of the moment is collected. The users can post on already-existing subreddits or they can create and moderate new ones on any topic of choice. In a subreddit, users can either create a new thread via a submission or indefinitely expand the conversation tree by commenting on an existing thread. The level of content moderation in a subreddit is solely decided by its moderators.}
    \label{fig:apx:1reddit}
\end{figure}
\newpage 
\begin{table}[h!]
    \small
    \centering
\begin{tabular}{l|ccccc}
\toprule
{Subreddits} & 2014 & 2015 & 2016 & 2017 & 2018 \\
\midrule
opiates               &             x &             x &             x &             x &             x \\
OpiatesRecovery       &             x &             x &             x &             x &             x \\
lean                  &             x &             x &             x &             x &             x \\
heroin                &             x &               &               &             x &             x \\
suboxone              &             x &             x &             x &             x &             x \\
PoppyTea              &             x &             x &             x &             x &             x \\
Methadone             &             x &             x &             x &             x &             x \\
Opiatewithdrawal      &             x &             x &             x &             x &             x \\
fentanyl              &               &             x &             x &             x &             x \\
codeine               &             x &             x &             x &             x &             x \\
HeroinHeroines        &               &               &               &               &             x \\
heroinaddiction       &               &             x &             x &             x &             x \\
oxycodone             &             x &             x &             x &             x &             x \\
opiatescirclejerk     &             x &             x &             x &             x &             x \\
loperamide            &               &               &             x &             x &             x \\
Opiate\_Withdrawal     &               &               &               &             x &             x \\
OpiateAddiction       &               &               &             x &             x &             x \\
PoppyTeaUniversity    &               &               &               &             x &             x \\
random\_acts\_of\_heroin &             x &             x &             x &             x &             x \\
Norco                 &               &               &             x &             x &             x \\
GetClean              &               &               &               &             x &             x \\
0piates               &             x &             x &             x &             x &             x \\
zubsolv               &             x &               &               &             x &             x \\
oxycontin             &             x &             x &             x &               &               \\
CodeineCowboys        &               &             x &             x &             x &               \\
OurOverUsedVeins      &             x &             x &             x &             x &             x \\
LeanSippersUnited     &               &               &               &             x &               \\
HopelessJunkies       &               &               &             x &             x &               \\
KetamineCuresOPIATES  &               &               &               &             x &               \\
AnarchyECP            &             x &               &             x &             x &               \\
PSTea                 &               &               &             x &               &               \\
glassine              &             x &             x &             x &             x &               \\

\bottomrule
\end{tabular}
\caption{Subreddits discussing firsthand nonmedical use of opioids. An X marks the presence of a subreddit in a specific year.}
\label{tab:apx:2subreddits}
\end{table}   
%
%
%%%%%%%%
\begin{table}[!htbp]
    \small
    \centering
    \begin{tabular}{cccccc}
    \toprule
         & Min term count  & Vector size & Context window & Negative Sampling & Training Epochs \\
    \midrule
     word2vec  & 5  & 256 & 5 & 10 & 200 \\
     GloVe    & 5  & 256 & 10 & - & 300\\
    \bottomrule
    \end{tabular}
    \caption{Relevant training parameters of the word embeddings. All the other parameters are set to default values. Two state-of-the-art word embedding models, namely word2vec, and GloVe, have been trained with all the comments and submissions in our subreddits dataset. After a-posteriori validation by a domain expert in terms of topical coherence, we choose word2vec as the reference word embedding model.}
    \label{tab:apx:3parameters}
\end{table}
\begin{figure}[!htbp]
    \centering
    \includegraphics[width=.9\textwidth]{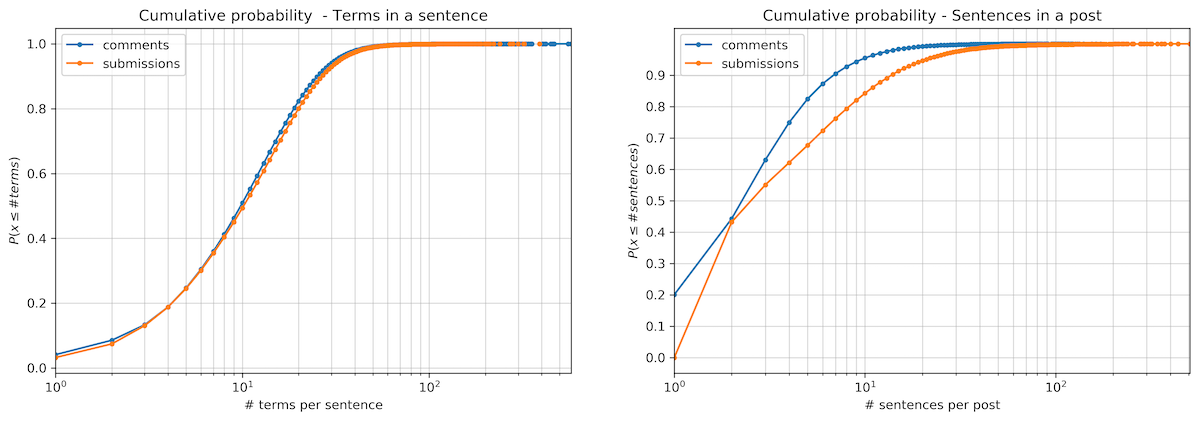}
    \caption{Cumulative probability of finding n or fewer terms in a sentence for submissions and comments (left panel). Cumulative probability of having n or fewer sentences in a submission or a comment (right panel). Plots refer to the selected subreddit in 2018.}
    \label{fig:apx:4cum}
\end{figure}
%
%
%
%%%%%%%%%%%%%%%%%%%%%%%%%%%%%%%%%%%%%%%%%%%%%
\begin{figure}[!htbp]
    \centering
    \includegraphics[height=.9\textheight]{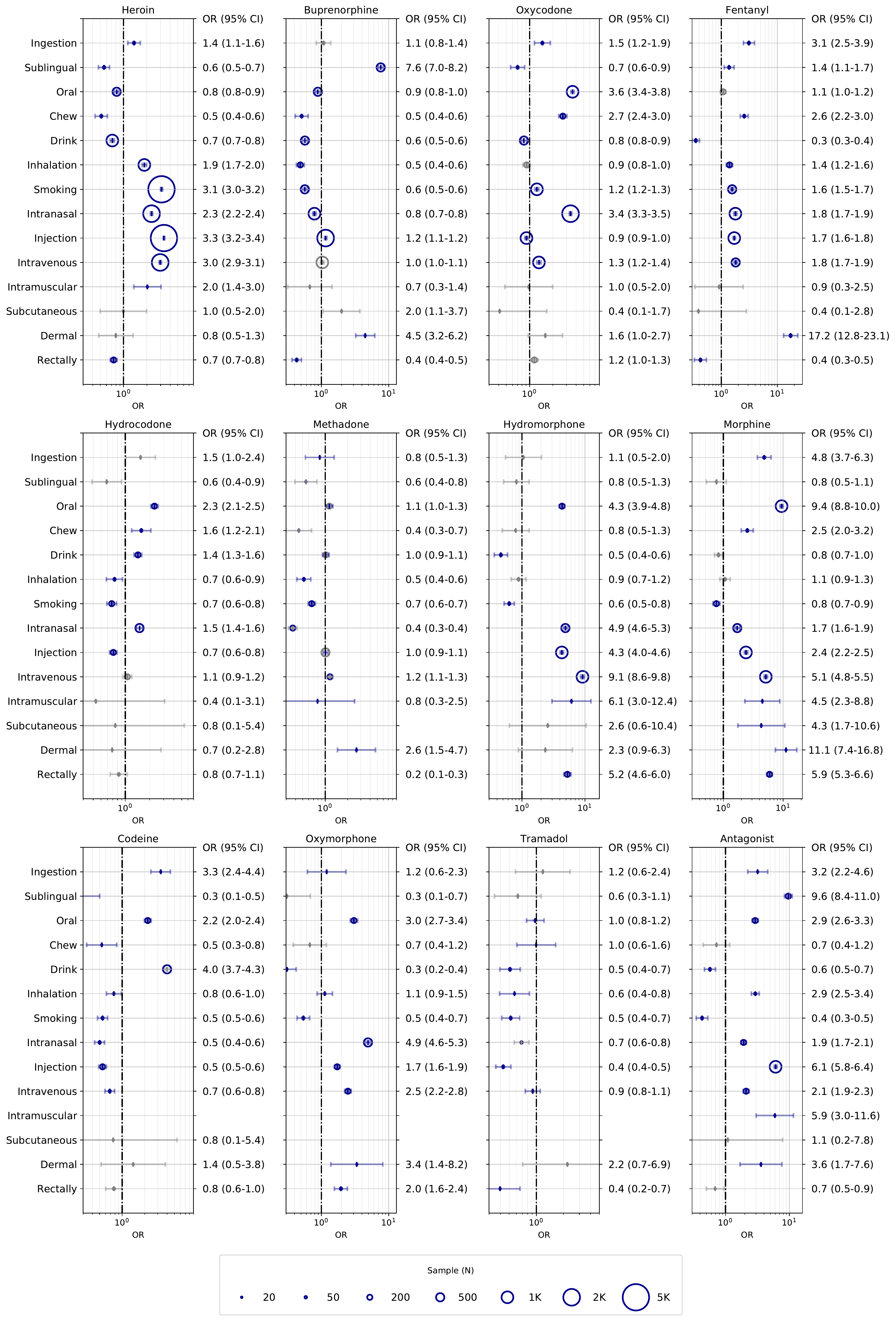}
    \caption{Odds Ratios of opioid substances and Secondary Routes of Administration. The central line and the bar mark the OR and the 95\% confidence interval respectively, while the size of the circle is proportional to the sample of co-mentions. Measures that are not statistically significant (P >.01) are reported in gray. Labels on the right axis report the Odds Ratio and the confidence interval.}
    \label{fig:apx:5ordrugsroa}
\end{figure}
 \begin{figure}[!htbp]
    \centering
    \includegraphics[height=.9\textheight]{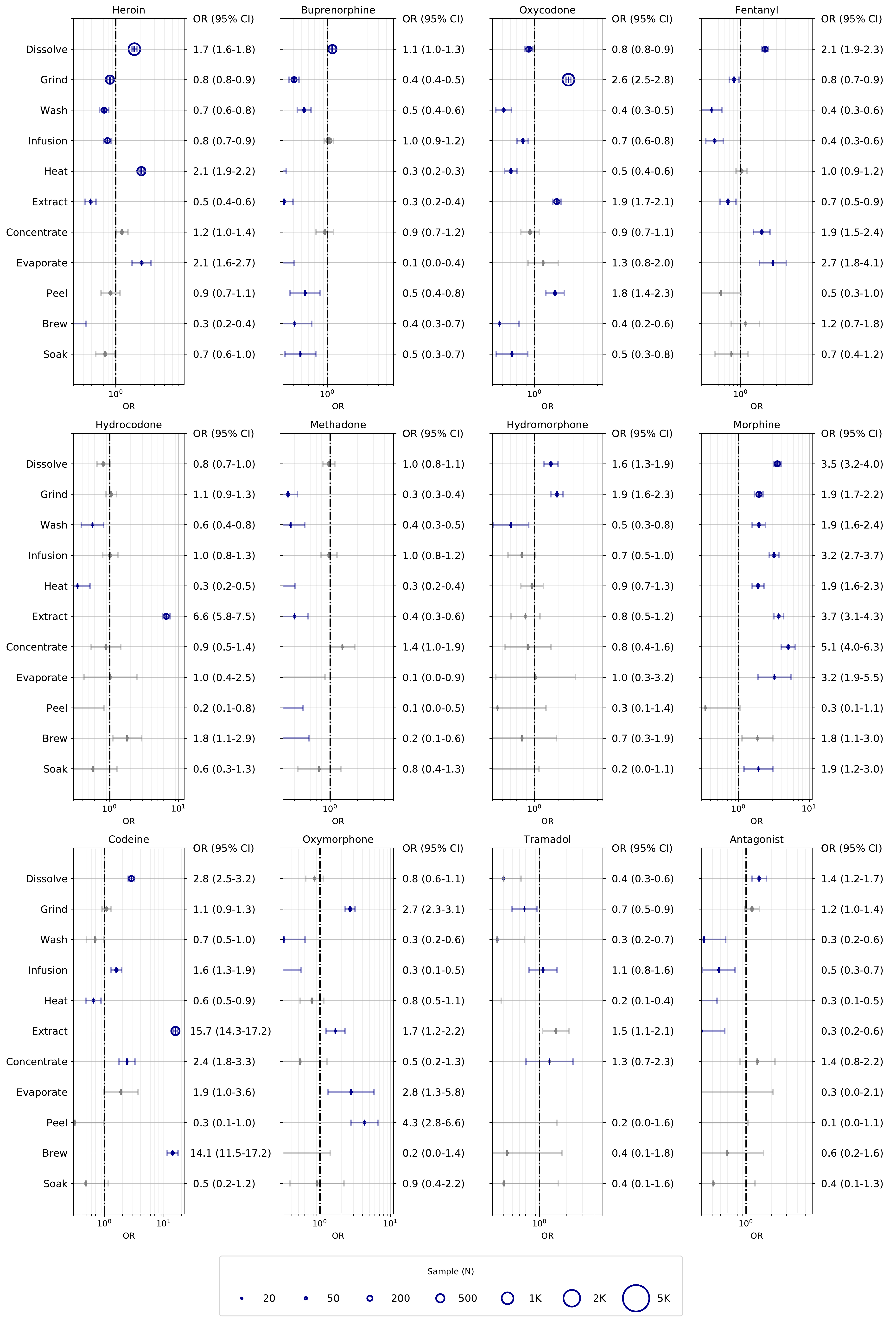}
    \caption{Odds Ratios of opioid substances and Drug Tampering Methods. The central line and the bar mark the OR and the 95\% confidence interval respectively, while the size of the circle is proportional to the sample of co-mentions. Measures that are not statistically significant (P >.01) are reported in gray. Labels on the right axis report the Odds Ratio and the confidence interval.}
    \label{fig:apx:6OR_drugs_trans_all}
\end{figure}
\begin{figure}[!htbp]
    \centering
    \includegraphics[height=.9\textheight]{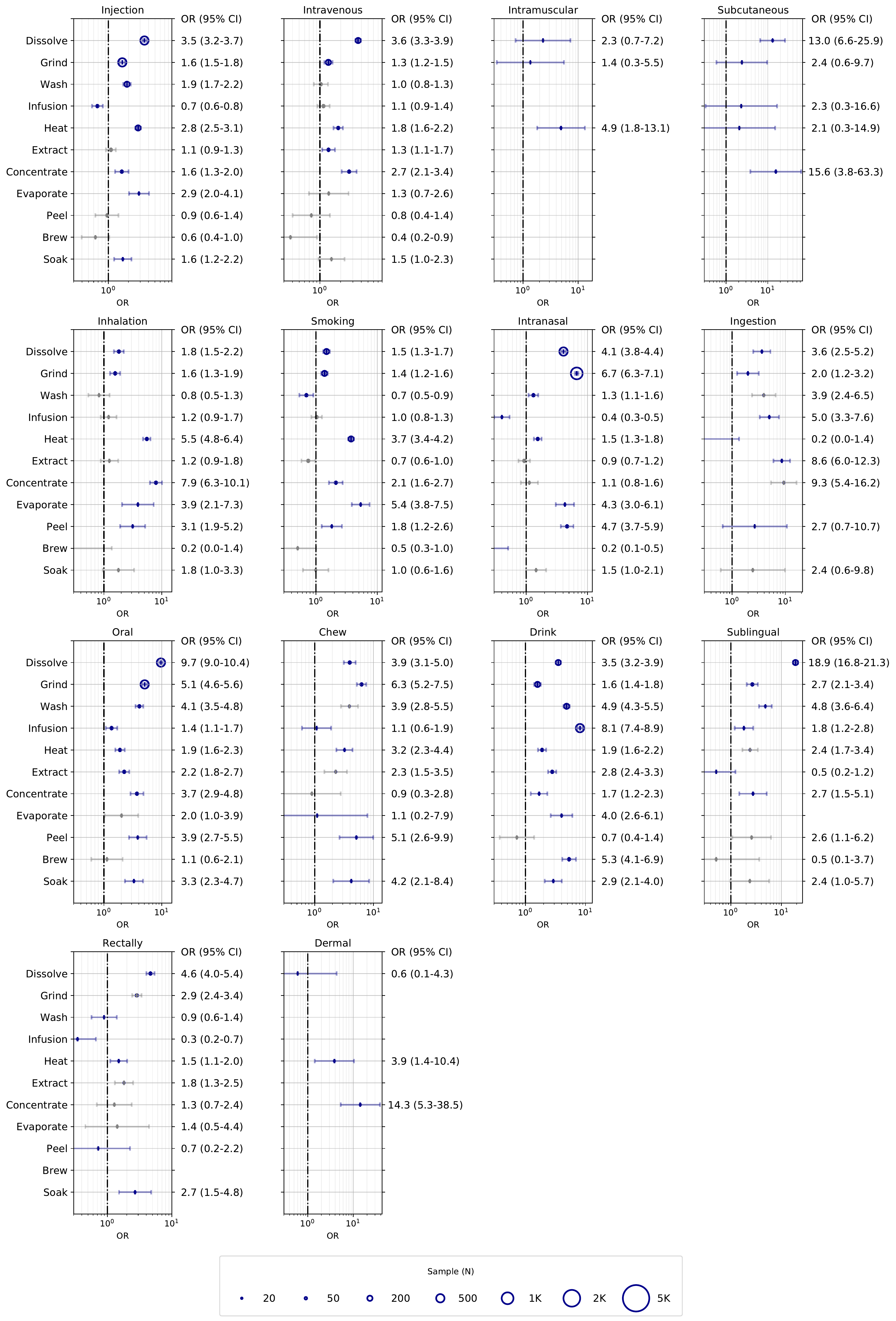}
    \caption{Odds Ratios of Secondary Routes of Administration and Drug Tampering Methods. The central line and the bar mark the OR and the 95\% confidence interval respectively, while the size of the circle is proportional to the sample of co-mentions. Measures that are not statistically significant (P >.01) are reported in gray. Labels on the right axis report the Odds Ratio and the confidence interval.}
    \label{fig:apx:7OR_trans_roa_all}
\end{figure}